\documentclass[aps,english,floatfix,twocolumn,10pt]{revtex4-2}
    \usepackage{graphicx}
    \usepackage[colorlinks=true,linkcolor=blue,citecolor=blue,urlcolor=blue]{hyperref}
    \usepackage{amsmath}
    \usepackage{amsfonts}
    \usepackage{orcidlink}
    \graphicspath{{Figs/}}
\begin{document}
        \title{Localization transitions in non-Hermitian quasiperiodic lattice}
 
	\author{Aruna Prasad Acharya\orcidlink{0000-0002-0404-9990}}
	\affiliation{Department of Physics and Astronomy, National Institute of Technology, Rourkela, Odisha-769008, India}
	
	\author{Sanjoy Datta\orcidlink{0000-0002-0308-9341}}
	\email[Corresponding author: ]{dattas@nitrkl.ac.in}
	\affiliation{Department of Physics and Astronomy, National Institute of Technology, Rourkela, Odisha-769008, India}
	
    \begin{abstract}
        
        The delocalization-localization (DL) transition in non-Hermitian systems exhibits intriguing features distinct from their Hermitian counterparts. In this study, we investigate the DL transition in a generalized non-Hermitian lattice with asymmetric hopping and complex quasi-periodic potential. Irrespective of the boundary conditions, the lattice undergoes a DL transition at a critical strength of the quasiperiodic potential with identical modulation of its real and complex parts. For periodic boundary conditions (PBC), we obtained an analytical expression that accurately predicts this critical point. Our numerical results indicate that the critical point remains the same with open boundary condition (OBC) as well. Interestingly, we observe that a difference in the modulation of the real and the complex part of potential leads to a mixed phase that appears between the delocalized and the localized phases. Intriguingly, within the mixed state region, we observed a coexistence of skin modes and localized states in the case of OBC, while in the case of PBC, a mixed phase is created by a coexistence of delocalized and localized states. We mapped out the phase diagrams for different scenarios offering valuable insights into the role of different parameters in a wide class of non-Hermitian quasiperiodic lattices.
    \end{abstract}
    \maketitle
    
    \section{Introduction}

        In condensed matter physics, Anderson localization is an active topic of discussion, although it fails to predict the localization transition for lower dimensions \cite{Anderson, Abrahams, Thomas_Guhr, B_Kramer}. It opens up the door to understanding the delocalization-localized (DL) transition. It fails the diffusion of the wave packets because of random disorder, leading to the localization of particles in lower dimensions. However, only the three-dimensional system shows a distinct metal-to-insulator phase transition around a critical point\cite{Abrahams}. Furthermore, the onsite quasidisorder system exemplified by the Aubry{\textendash}Andr{\'{e}}-Harper (AAH) model \cite{Aubry, Harper} has an identical phenomenon where the potential strength controls DL transition even for the one-dimensional system.

        Due to its exotic properties, numerous extensions of the AAH model in many ways have been explored, such as modified hopping to an exponential decay (short-range), long-range hopping \cite{Biddle-2010}, and modulated hopping amplitude with some irrational frequency, to name a few. They show some exciting phenomena in DL transition, like mobility edges, multifractals, etc. This leads to growing interest in this system and various extensions work \cite{Madsen, Lang, Vidal, Kraus-106402, Ganeshan}. Furthermore, many theories and experimental realizations have been made in photonic \cite{Lahini, Verbin, Kraus, Dal_Negro} and ultracold atomic systems \cite{Teng_Xiao, Modugno_2010}.
        
        \begin{figure*}
            \centering
            
            \includegraphics[scale=0.65]{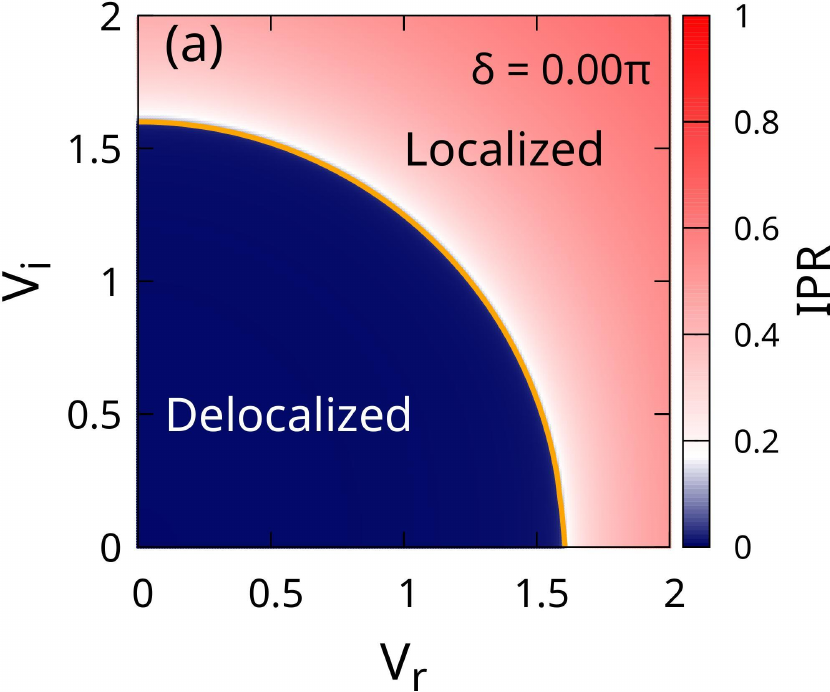}\hspace{4mm}
            \includegraphics[scale=0.65]{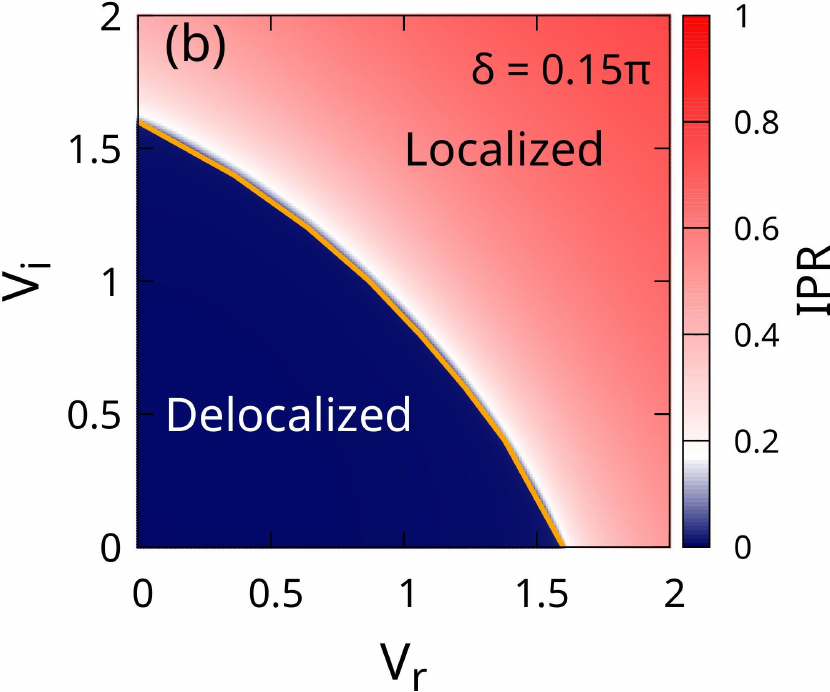}\hspace{4mm}
            \includegraphics[scale=0.65]{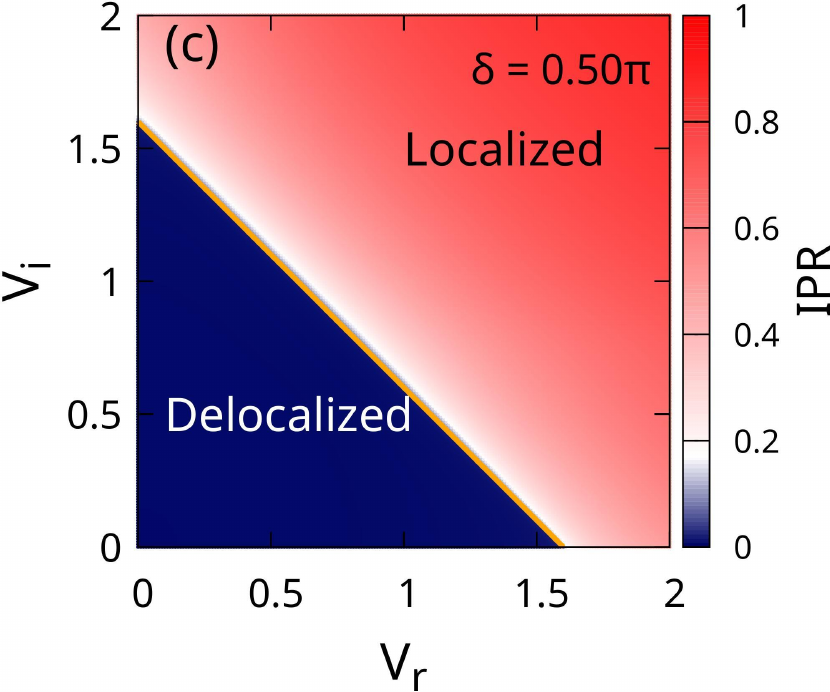}
        
            \caption{The above depict phase diagrams are for a system with L=610 and asymmetric hopping strengths $t_l=0.60$ and $t_r=0.80$. The offset phases $\delta$ in Figures (a)-(c) are 0, 0.15$\pi$, and 0.50$\pi$, respectively. The bold orange line separates the delocalized (\emph{dark blue}) and localized (\emph{red}) states, representing the analytically predicted critical values. The $\rm IPR_{avg}$ is used to distinguish the two phases in all figures.
            }
            \label{Fig: asymmetric phase space}
        \end{figure*}        

        The growing interest in the non-Hermitian systems is driven by the possibility of a wide range of exciting physical phenomena, which are not possible in the Hermitian counterparts. For example, more than two decades ago, it was demonstrated \cite{HatanoNelson1996, HatanoNelson1998} that a 1D lattice with random onsite potential and asymmetric hopping amplitude undergoes a DL transition at a critical disorder strength. This DL transition is accompanied by a complex-real transition of the eigenspectrum at the same critical point. This paradigmatic non-Hermitian Hamiltonian, where the non-hermiticity is introduced via the asymmetry in the hopping amplitude, is referred as the Hatano-Nelson (HN) model. More recently, it has been shown \cite{Jiang} that replacing the random potential in the HN model with a quasi-periodic potential does not affect the qualitative features of the DL transition. There exists another class of non-Hermitian models, where the non-hermiticity is introduced by considering a complex potential while the hopping remains symmetric. For example, in a non-Hermitian version of the 1D AAH model \cite{Longhi-2019, Longhi} consisting of potential with complex QP phase, by analytical computation of the Lyapunov exponent (LE), it is demonstrated that a similar DL transition takes place. 

        However, in contrast to the HN type models, this DL transition is accompanied by a real-complex transition of the eigenspectrum concomitant with the breaking of $\mathcal{PT}$ symmetry. Long before this observation, the DL transition and its relationship with the eigenspectrum were studied numerically \cite{Jazaeri} in 1D with a more generalized form of the complex QP potential. In this work, the strengths of the real and the complex parts of the QP potential are generally unequal. Recently, the critical properties of the model Hamiltonian of Ref.~\cite{Cai_2022} have been studied in the presence of $p$-wave pairing. Apart from the $p$-wave pairing, there is a phase difference between the real and the complex parts of the QP potential.
        
        In another recent research, Y.Liu et al. studied \cite{Liu2021} the interplay of the localization transition and the eigenspectrum in a 1D non-Hermitian lattice in the presence of nonreciprocal hopping and complex QP potential.
        
        Motivated by these results, in this present work, we have studied a generalized 1D non-Hermitian lattice with asymmetric hopping and complex QP potential. To generalize our model Hamiltonian, along with the asymmetric hopping we have considered a real and complex QP potential whose strengths are unequal in general. Furthermore, we have considered an offset phase between these two QP potentials. By controlling the offset phase and the amount of asymmetry in the hopping, different limiting cases that have been discussed earlier can be reproduced. To determine the DL transition analytically, we have computed the LE following Ref.~\cite{Liu2021}. Our analytical results suggest a DL transition at a critical point, and the shape of the phase boundary crucially depends on the offset phase for a fixed value of the asymmetry in the hoppings. Moreover, for a fixed value of the offset phase, the critical points depend on an effective value of the QP potential and the amount of asymmetry in the hoppings. The offset phase dependence of the phase boundary remains qualitatively similar with respect to a change in the value in hopping asymmetry. Our analytical results have been verified numerically. Furthermore, we have numerically verified that a change from periodic to open boundary conditions does not alter the phase boundary. However, with boundary conditions, we have observed the skin effect in the delocalized phase. It is important to note that the spectrum remains completely complex across the critical point irrespective of the boundary conditions, owing to the fact that $\mathcal{PT}$ symmetry is broken due to the non-reciprocity of the hoppings.

        The discussion in this paper is arranged as follows. In Sec: \ref{Sec: Model}, a generalized non-Hermitian version is discussed with two QP potentials. Methods to understand the phase transition point numerically is discussed in Sec: \ref{Sec: Numerical methods}. Then the Sec: \ref{Sec: Result and discussion} is divided into two subsections, where DL transition in Periodic Boundary Condition (PBC) system is discussed in Sec: \ref{Sec: PBC} with analytical and numerical results and followed by Open Boundary Condition (OBC) system in Sec: \ref{Sec: OBC}. A different situation where the frequencies of two QP potentials are unequal, showing mixed states, has been discussed in Sec.\ref{Sec: Different Freq}.  Finally, the whole summary of this work is discussed in Sec: \ref{Sec: Concusion}.
        
    \section{Model}\label{Sec: Model}
        
        We start with a 1D quasiperiodic lattice model with a complex potential and asymmetric hopping amplitudes. The Hamiltonian is given as,  

        \begin{eqnarray}
            E \psi_n = (t_r \psi_{n+1} + t_l \psi_{n-1} )+ V_n \psi_n,
            \label{Eq: Dual_QP_Hamiltonian}
        \end{eqnarray}
        $t_l(t_r)$ is the hopping amplitude to the left (right), and $n$ represents the site index. The on-site potential $V_n$ consists of real and complex quasiperiodic potentials, and it is given by
        \begin{eqnarray}
            V_n = V_r \cos(2\pi{\beta}\textit{n} + \phi) + i V_i \cos(2\pi{\beta}\textit{n} + \phi +\delta).
            \label{Eq: Quasi Potential}
        \end{eqnarray}
        
        Here, $V_r(V_i)$ is the strength of the real(non-real/complex) part of the potential. The quasi-periodic modulation of the underlying potential is due to the parameter $\beta= (\sqrt{5}-1) / 2$, which is the inverse of the golden ratio. The modulation of the potential could be approximated as $\lim_{n \to \infty} F_{n-1}/F_n$ with $F_n$ and $F_{n-1}$ being two consecutive numbers in the Fibonacci series. The lattice system size is set to $L=F_n a$ (lattice constant $a$ = 1 in arb. units) to preserve the QP in the potential. The global phase $\phi$ shifts the two QP potentials spatially, while $\delta$ measures the phase difference between them. Here, we have set $\phi = 0$ (unless mentioned otherwise) in this discussion. The fundamental difference between the Hamiltonian model considered in this work and the previous works is that $V_r \neq V_i$ in general, and there is a phase difference $\delta$ between the real and the complex part of the QP potential in the presence of asymmetric hoppings. As mentioned in the introduction, various non-Hermitian QP Hamiltonians have already been studied in the literature. As we are going to see that the results of these studies can be reproduced from our results as limiting cases. For example, by setting $V_i=0$ and $t=t_l=t_r$, we can revisit the classic Hermitian AAH model \cite{Aubry},  which has a DL transition at $2t=V_r$. Moreover, setting $t_l \not = t_r$, we obtain the Hatano-Nelson Hamiltonian \cite{HatanoNelson1996, HatanoNelson1998} with QP potential. The DL transition, in this case, is given by the condition $2max(t_l,t_r)=V_r$ \cite{Jiang}. Considering $V_i \neq 0$ and $t=t_l=t_r$, we obtain the previously studied non-Hermtian models of Ref.~\cite{Jazaeri} ($\delta=\pi/2$) and Ref.~\cite{Cai_2022} in the limit of vanishing p-wave pairing amplitude. Some other classes of non-Hermitian AAH models \cite{Longhi, Cai-Jan-2021}, where a complex phase is added to the potential ($V \cos(2\pi\beta n +\phi +ih)$), can easily be obtained from our Hamiltonian by setting  $V_r = V \cosh(h)$, $V_i= V \sinh(h)$, $\delta=-\pi/2$, and appropriate hopping amplitudes. Similarly, with $V_{r/i}=V$, $\delta=\pi /2$, $\phi=0$, and appropriate hopping strengths we obtain the results of Refs.~\cite{Longhi-2019} and \cite{Liu-2021}. 
        
        In this entire discussion, we have chosen $t=1$ (unless otherwise specified), and all quantities are scaled accordingly.

    \section{Analytical determination of the phase boundary}\label{App: LE_II}
            
        The widely used approach to determine the critical point of DL transition in such QP systems is to compute the LE. In the delocalized phase LE is zero, while it is a non-zero positive quantity in the localized phase. Recently, the analytical expression of the LE has been obtained for the limiting case of symmetric hopping in Ref.~\cite{Cai_2022}, which agrees  with the numerical results obtained in Ref.~\cite{Jazaeri} 
        (with $\delta=\pi/2$ and vanishing p-wave pairing amplitude). 
        In this work, we have analytically computed the LE for the case of asymmetric hopping following Ref.~\cite{Cai_2022}, and obtained the phase diagram in the parameter space of $V_r$ of $V_i$. It is important to note that the qualitative nature of the phase diagram turns out to be similar to the symmetric hopping case in the limit of vanishing p-wave pairing amplitude \cite{Cai_2022}.
        
        For the ease of discussion, we set $t_r>t_l$, where $t_r=t'e^{\eta}$ and $t_l=t'e^{-\eta}$ is being set in our Hamiltonian (Eq. \ref{Eq: Dual_QP_Hamiltonian}). Now, our Hamiltonian could be written as,
        \begin{eqnarray}
            t'(e^{\eta} \psi_{n+1} +e^{-\eta} \psi_{n-1}) + V_r  \cos(2\pi{\beta}\textit{n} + \phi) \psi_n  \nonumber  & &\\
            + i V_i \cos(2\pi{\beta}\textit{n} + \phi +\delta)\psi_n = E \psi_n ,
        \label{Eq: Model-II}
        \end{eqnarray}
        
        where $t'=\sqrt{t_r t_l}$ and $\eta=\ln \big(\sqrt{t_r/t_l}\big)$. To decide the phase boundary, the LE can be calculated with the help of the method proposed by Thouless \cite{Thouless}, which gives the LE in terms of an integral involving the density of states. In this approach, the LE corresponding to a single-particle Hamiltonian $H$ is given by
        
        \begin{eqnarray}
            \gamma= \int d\epsilon \rho(\epsilon) \ln|\epsilon-E_B|-\ln(J),
            \label{EQ: LE_2}
        \end{eqnarray}
        
        where $E_B$ is a reference energy, $\epsilon$ represents the eigenvalues of the Hamiltonian, and the density of the states is given by $\rho(\epsilon)$. The additional term $\ln J$ is introduced to reset the energy scale, which will be a crucial component to compute the LE in the case of asymmetric hopping. In the case of symmetric hopping $J$ is equal to the hopping amplitude \cite{Cai_2022}. Furthermore,  we set $E_B=0$ for simplicity since it is found numerically that LE is independent of the choice of $E_B$.  
        For the computation of LE, it is convenient to write $\gamma$ as 
        
        \begin{equation}
         \gamma = g/L - \ln(J),
         \label{Eq: gamma}
        \end{equation}
        where $g$ is defined as 
        \begin{eqnarray}
            g\equiv \ln|det(H)| \equiv \sum^{L}_{n=1} \ln |\epsilon_n|.
            \label{EQ: g}
        \end{eqnarray}
        In the above equation $|\epsilon_n|$ represents the n-th eigenvalue of the 
        Hamiltonian $H$ of Eq. \ref{Eq: Model-II}, which reads as
        \begin{equation}
	       H=
		\begin{pmatrix}
			V_1 & t'e^\eta & 0 & ... & 0 & 0 & t'e^{-\eta}  \\
			t'e^{-\eta} & V_2 & t'e^\eta & ... & 0 & 0 & 0 \\
			0 & t'e^{-\eta} & V_3 & ... & 0 & 0 & 0 \\
			... & ... & ... & ... & ... & ... & ... \\
			0 & 0 & 0 & ... & t'e^{-\eta} & V_{L-1} & t'e^\eta \\
			t'e^\eta & 0 & 0 & ... & 0 & t'e^{-\eta} & V_L \\
		\end{pmatrix}
        \label{Eq:Hamiltonian_matrix_H}
        \end{equation}

        The summation in Eq.~\ref{EQ: g} can be replaced with integration in the limit of  large L (i.e., $L \to \infty$),
        
        \begin{eqnarray}
            g \equiv L \int d\epsilon \rho(\epsilon) \ln|\epsilon|.
        \end{eqnarray}

        To calculate $det(H)$, we first perform a similarity transformation $SHS^{-1}$ with $S=diag(e^\eta,e^{2\eta},...,e^{L\eta})$. $SHS^{-1}$ reads as,
        
        \begin{equation}
		SHS^{-1}=
		\begin{pmatrix}
			V_1 & t' & 0 & ... & 0 & 0 & t'e^{L\eta}  \\
			t' & V_2 & t' & ... & 0 & 0 & 0 \\
			0 & t' & V_3 & ... & 0 & 0 & 0 \\
			... & ... & ... & ... & ... & ... & ... \\
			0 & 0 & 0 & ... & t' & V_{L-1} & t' \\
			t'e^{-L\eta} & 0 & 0 & ... & 0 & t' & V_L \\
		\end{pmatrix}
        \label{Eq: similarity transform matrix H1}
        \end{equation}
        
        It is clear that in the large L limit ($, i.e., L \to \infty)$, the element $(H)_{L,1} \to 0$. Hence, $det(H)$ can now be written as,
        
        \begin{eqnarray}
            det(H) = (-1)^{L+1}t'^Le^{L\eta}+det(H_1),
            \label{EQ: Det_H_PBC}
        \end{eqnarray}
        where $H_1$ is given by
        \begin{eqnarray}
             H_1=
		\begin{pmatrix}
			V_1 & t' & 0 & ... & 0 & 0 & 0  \\
			t' & V_2 & t' & ... & 0 & 0 & 0 \\
			0 & t' & V_3 & ... & 0 & 0 & 0 \\
			... & ... & ... & ... & ... & ... & ... \\
			0 & 0 & 0 & ... & t' & V_{L-1} & t' \\
			0 & 0 & 0 & ... & 0 & t' & V_L \\
		\end{pmatrix}
        \end{eqnarray}
        
        To determine $det(H_1)$, we exploit the relation between the determinant of a Hamiltonian and the corresponding LE as given in Eq.~\ref{Eq: gamma}. Instead of computing the determinant directly, we first compute the LE for the Hamiltonian $H_1$ (see appendix \ref{App: LE_I}). The LE corresponding to $H_1$ is given by
        
        \begin{align}
            \gamma_1 =max \left\{\ln \frac{V_{eff}}{2t'},0\right\},
            \label{Eq: gamma1}
        \end{align}
        
        where $V_{eff}=\sqrt{V_r^2 + V_i^2 + 2 V_i V_r|\sin \delta|}$. Now, using Thouless's result (Eq. \ref{Eq: gamma}), the LE for the Hamiltonian $H_1$ can be written as
        
        \begin{equation}
            \gamma_1 = \frac{\ln|det(H_1)|}{L} - \ln(t').
        \end{equation}
        
        Finally, using Eq.~\ref{Eq: gamma1}, $det(H_1)$ can be written as,
        
        \begin{eqnarray}
            \lim\limits_{L \to \infty} det(H_1)  = \zeta~ max \big\{ t'^L,(V_{eff}/2)^L \big\},
            \label{EQ: det_H_OBC}
        \end{eqnarray}
        
        where $\zeta$ takes care of the possible overall sign.After substituting Eq.~\ref{EQ: det_H_OBC} in Eq.~\ref{EQ: Det_H_PBC}, we have
        
        \begin{eqnarray}
            \begin{aligned}
                \lim\limits_{L \to \infty}det(H) &= (-1)^{L+1}t'^Le^{L\eta} + \zeta ~max \left[ t'^L,(V_{eff}/2)^L  \right] \\
                & =\zeta'~ \big[ max \big\{ {(2t'e^\eta)},V_{eff} \big\} \big]^L,
            \end{aligned}
        \end{eqnarray}
        
        Here, $\zeta$ and $\zeta'$ are possible signs. The calculation of LE now becomes
        
        \begin{eqnarray}
            \begin{aligned}
                 \gamma &= \ln \Big\vert max \left[ {2t'e^\eta},V_{eff} \right] / J \Big\vert \\
                 &= 
                 \begin{cases}
                        \ln |2t'e^\eta/J|, &\mbox{if } ~2t'e^\eta > V_{eff}, \\
                        \ln |V_{eff}/J|, & \mbox{if } ~2t'e^\eta < V_{eff}. 
                \end{cases} 
            \end{aligned}
        \end{eqnarray}
        
        To determine the energy scale $J$, we note that in the absence of $V_{eff}$ the LE must be zero. This leads to the condition $J=2t'e^\eta$. Once $J$ is determined the LE in the presence of $V_{eff}$ for the QP Hamiltonian with asymmetric hoppings can be written as, 
        
        \begin{eqnarray}
            \gamma= max \bigg\{ \frac{1}{2}\ln \frac{V_r^2 + V_i^2 + 2 V_i V_r|\sin \delta|}{(2t'e^\eta)^2} ,0 \bigg\}.
            \label{EQ: LE_asymmetry}
        \end{eqnarray}
        
        The relation determines the critical point for the DL transition
        
        \begin{eqnarray}
                V_r^2 + V_i^2 + 2 V_i V_r|\sin \delta|&=(2t'e^\eta)^2 = (2 t_r)^2
                \label{Eq: Critical}
        \end{eqnarray}
        
        Similarly, for the case when $t_r<t_l$, we could decide the critical point of DL transition by just setting $-\eta$ in all the analytical calculations above. It will read as,
        
        \begin{eqnarray}
                V_r^2 + V_i^2 + 2 V_i V_r|\sin \delta|&=(2t'e^{-\eta})^2 = (2 t_l)^2
        \end{eqnarray}
        
        On the basis of these two results, a more generalized version of the DL transition point could have been written as,
        
        \begin{equation}
            V_r^2 + V_i^2 + 2 V_i V_r|\sin \delta| = \big( 2 max \big\{ t_l,t_r \big\} \big)^2
            \label{Eq: DL critical point}
        \end{equation}
        
        It is easy to see that in the limiting case of $\delta=0$, the phase boundary is a circle, while in the other extreme limit of $\delta=\pi/2$, it follows an equation of a straight line. For any other value of $\delta$, the phase boundary lies between these two extreme limits. Furthermore, by setting $V_i=0$ (or $V_r=0$) in Eq. \ref{EQ: LE_asymmetry}, the LE for real (or purely complex) QP potential with asymmetric hopping is obtained as,
        
        \begin{eqnarray}
            \gamma_{r(i)}=max\{\ln|V_{r(i)}/2t'e^\eta|,0\}.
        \end{eqnarray}
        
        The critical point, in this case, is given by the condition 
        
        $V_{r(i)} = 2t'e^\eta.$
        
    \section{Numerical methods} \label{Sec: Numerical methods}
        In this section, we introduce the numerical tools used to determine the critical point and the nature of the eigenspectrum. To locate the DL transition, we proceed with one of the most familiar indicators, which is the average inverse participation ratio ($\rm{IPR}_{avg}$) \cite{Mirlin, Wessel}, which reads as,
        \begin{equation}
            \rm{IPR}_{avg}=\frac{1}{L}\sum_i \rm{IPR}^{(i)},
            \label{Eq: IPR_avg}
        \end{equation}
        where  $\rm{IPR}^{(i)}$ is given by
        \begin{equation}
	    \rm{IPR}^{(i)}=\sum_n|\psi_{n}^i|^4 .
            \label{Eq: IPR}
        \end{equation}
        In the above expression, $\psi_{n}^i$ is the normalized eigenstate corresponding to the $i$-th eigenvalue, and $n$ is the site index.
            For a completely localized state IPR $\simeq 1$, while it scales inversely with the system size ($\simeq 1/L$) when an eigenstate is  delocalized.
            
        As discussed in the introduction, typically in non-Hermitian systems, the DL transition is accompanied by a drastic change in the nature of the eigenspectrum. To capture the possible change of the eigenspectrum at the critical point, we compute the density of states (DOS) of complex energies and the maximum value of the imaginary $E$, which is defined as,
        \begin{eqnarray}
	        \rm{DOS} = \frac{N(Img(E) \neq 0)}{L} ,~~~~~~~~~~~~ 
                \label{Eq: max_E}
        \end{eqnarray}
        where, $N(Img(E)\neq 0)~$ is the number of eigenstates with non-zero complex eigenenergies.
        \begin{eqnarray}
            max (Img (E))=\max\limits_{i \in \{1,.....L\}}(Img (E_i)).        
            \label{Eq: DOS}
        \end{eqnarray} 
                        
        \begin{figure*}
            \centering
            \includegraphics[scale=0.7]{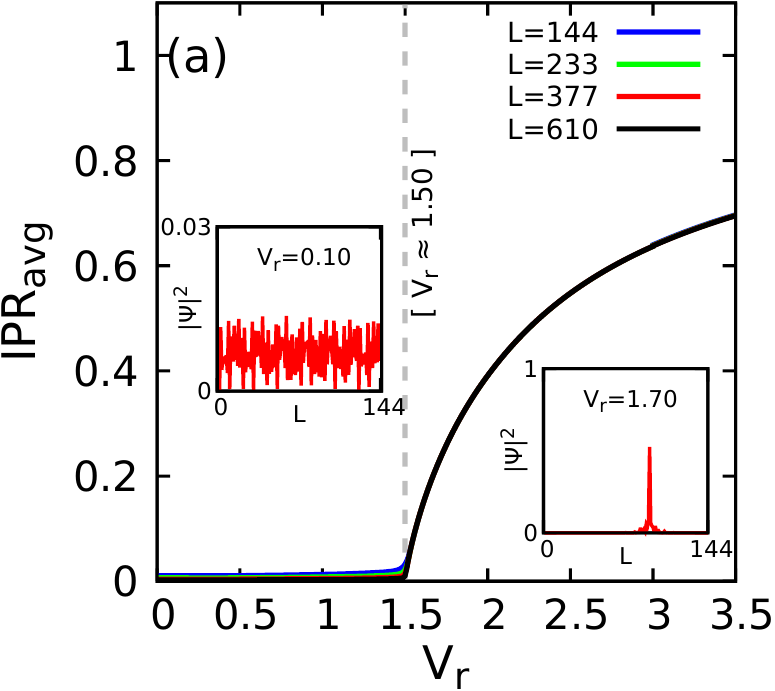}\hspace{8mm}
            \includegraphics[scale=0.7]{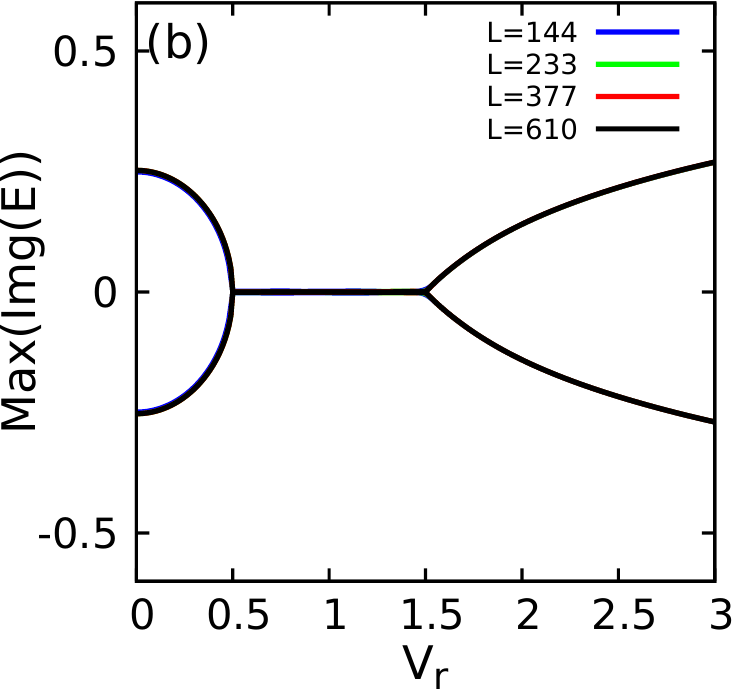}\hspace{8mm}
            \includegraphics[scale=0.7]{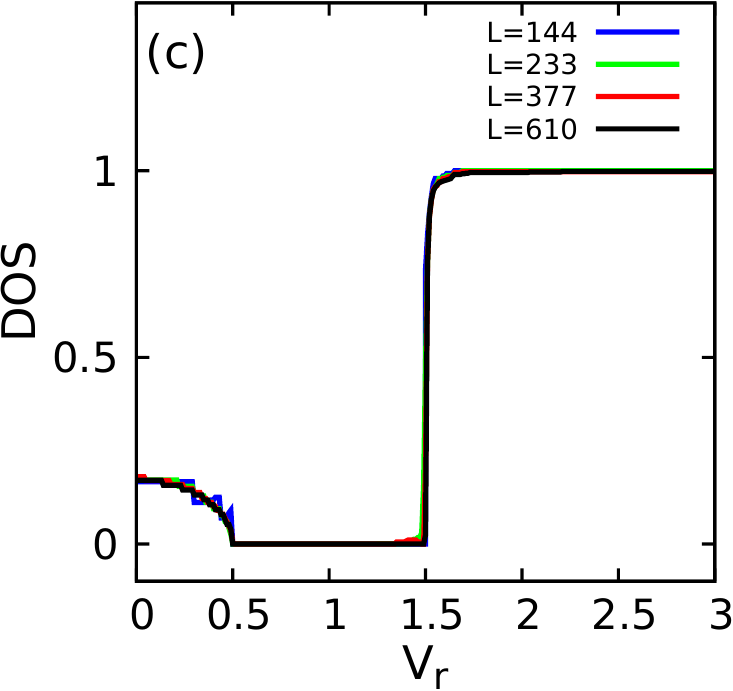} 
            \caption{The figures displayed above correspond to different system sizes: L=144, 233, 377, and 610. The amplitude of the imaginary QP potential is set to $V_i=0.50$, and the hoppings are symmetric, with $t=t_l=t_r=1.00$. The offset phase is set to $\delta=0.50\pi$.}
            \label{Fig: symmetric IPR E DOS}
        \end{figure*}
            
        \begin{figure*}
            \centering
            \includegraphics[scale=0.7]{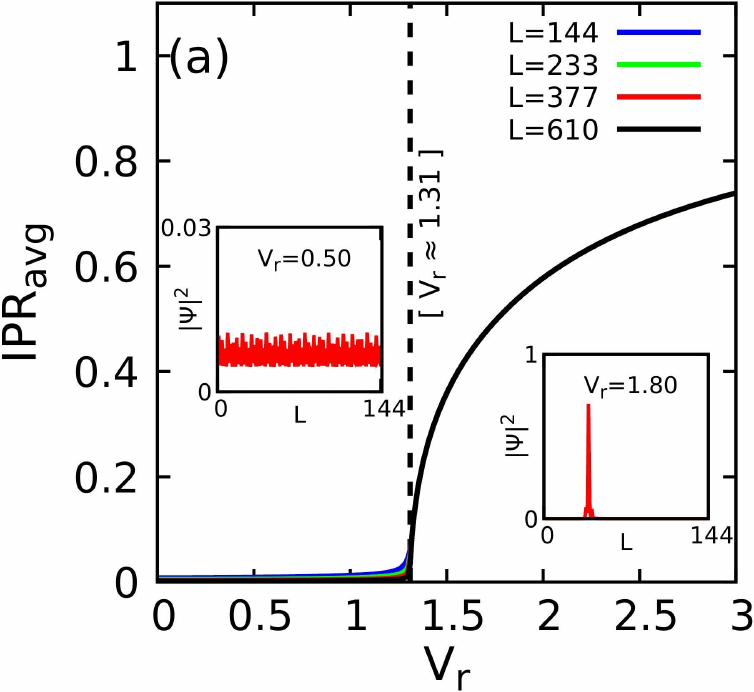}\hspace{8mm}
            \includegraphics[scale=0.7]{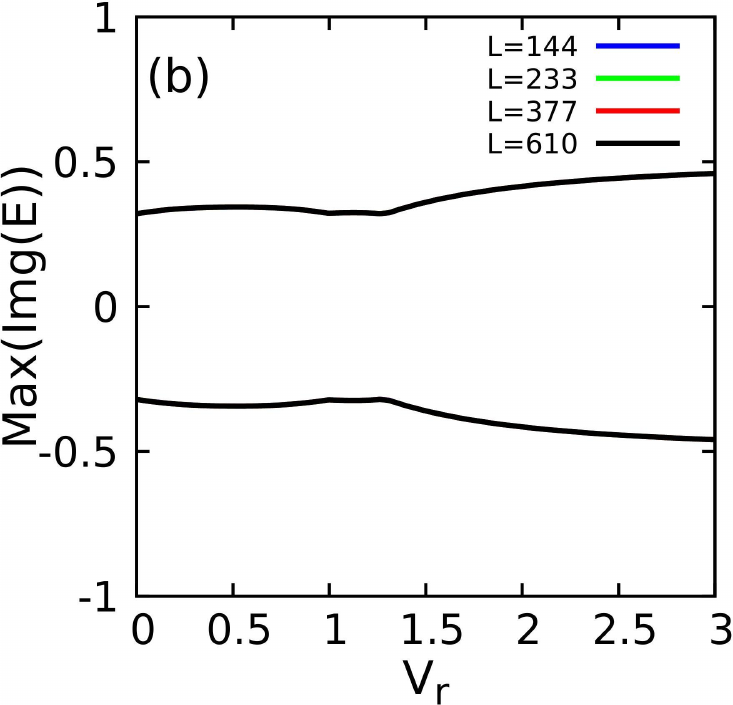}\hspace{8mm}
            \includegraphics[scale=0.7]{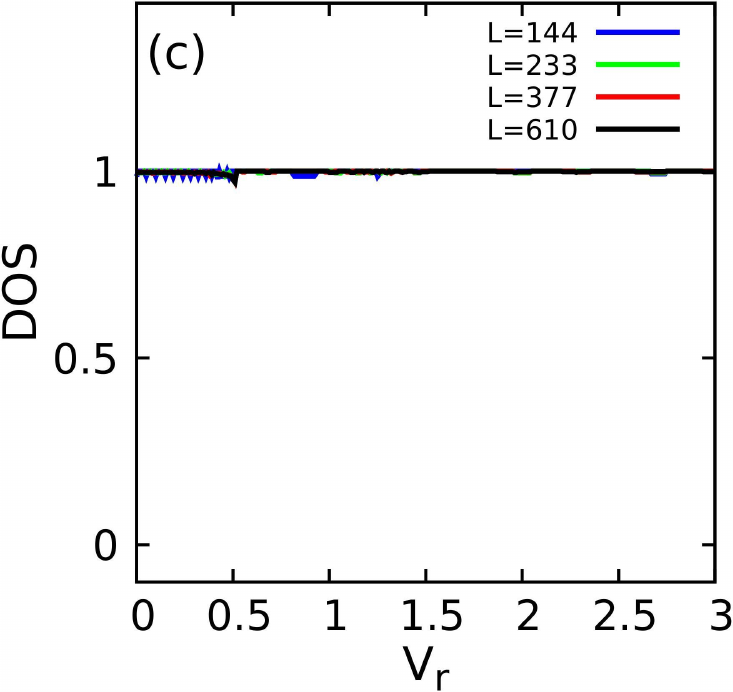}
            \caption{The figures displayed above illustrate for various system sizes, namely L=144, 233, 377, and 610. These diagrams depict the effects of an asymmetric configuration, where the hopping parameters have different values: $t_l=0.60$ and $t_r=0.80$. The offset phase is set to $\delta=0.15\pi$. Additionally, the amplitude of the imaginary QP potential is kept constant at $V_i=0.50$.}
            \label{Fig: asymmetric IPR E DOS}
        \end{figure*} 

    \section{Results and discussion}\label{Sec: Result and discussion}
    
        We now present our results. In Sec.~\ref{Sec: PBC}, we first discuss the results of a 1D lattice with PBC, while the corresponding results with OBC are presented in Sec.~\ref{Sec: OBC}. The results for the situation where the modulating frequencies of the two QP potentials are unequal have been discussed in Sec.\ref{Sec: Different Freq} 

        \subsection{DL phase and eigenspectrum transition Under PBC}\label{Sec: PBC}

            In Fig.~\ref{Fig: asymmetric phase space}, we have presented one of the main findings of our work. The numerically obtained $\rm{IPR}_{avg}$ are presented as a function of the parameters $V_r$ and $V_i$ along with the analytically predicted critical point given by Eq.~\ref{Eq: DL critical point}. From these phase diagrams, it is clear that when the modulating frequencies of the two QP potentials are equal, there exists a phase boundary that separates delocalized and localized regimes in the presence of asymmetrical hoppings. The numerically obtained phase boundaries agree excellently with our analytical results. The two phases are represented with two different colors; $dark-blue$ indicates the delocalized phase, and the localized phase is labeled with $red$ color. The bold $orange$ line represents the analytically predicted phase boundary given in Eq.~\ref{Eq: DL critical point}. Qualitatively similar observations have been made in the case of symmetric hoppings in Ref.~\cite{Cai_2022}.
            
            To compute $\rm{IPR}_{avg}$, we have considered a lattice with $L=610$ with PBC. After numerically diagonalizing the Hamiltonian (Eq. \ref{Eq: Dual_QP_Hamiltonian}), we use the right eigenvectors to calculate the
            $\rm{IPR}_{avg}$.

            Fig. \ref{Fig: asymmetric phase space}(a) represents the phase-diagram for QP potentials with an offset phase $\delta=0.00$ (in-phase), and hopping amplitudes $t_l=0.60$ and $t_r=0.80$. In this case, numerically obtained critical points follow a quarter circular path whose radius agrees excellently with the analytically obtained result in Eq.~\ref{Eq: DL critical point}. The radius is given by $2t'e^\eta \approx 1.60$. The shape of the phase boundary crucially depends on the offset phase $\delta$. This is easily visible from the phase diagrams presented in Fig. \ref{Fig: asymmetric phase space}(b) and (c). In these figures, the offset phase is set to the value of  $\delta=0.15\pi$ and $\delta=0.50\pi$ respectively while keeping the strengths of the hoppings the same as in Fig.~\ref{Fig: asymmetric phase space}(a).
            
            Even though the qualitative nature of the phase diagram remains the same for symmetric and asymmetric hopping amplitudes, there are crucial differences in the energy spectrum. To illustrate the similarities and the differences between these two cases, we have compared the $\rm{IPR}_{avg}$, the energy spectrum, and the DOS in Figs.~\ref{Fig: symmetric IPR E DOS} and \ref{Fig: asymmetric IPR E DOS}. In the case of symmetric hoppings ($t_l=t_r=t$), the Hamiltonian is $\mathcal{PT}$ symmetric only when the offset phase $\delta = \pi/2$, while it is absent in the Hamiltonian with asymmetric hoppings irrespective of the value of the offset phase. From Figs.~\ref{Fig: symmetric IPR E DOS}(a) and \ref{Fig: asymmetric IPR E DOS}(a) it is evident that apart from the exact location of the critical point, the qualitative nature of the DL transitions is identical for symmetric and asymmetric hoppings. However, as a consequence of the $\mathcal{PT}$ symmetry, with symmetric hoppings, the energy spectrum becomes either partially or completely real in the delocalized regime. Interestingly, the spectrum becomes entirely real when $V_r > V_i$ \cite{Jazaeri}. This can be seen from Fig~\ref{Fig: symmetric IPR E DOS}(b). However, the spectrum becomes purely complex along with the transition to the localized phase. This transition of the energy spectrum to purely complex at the critical point is evident from the DOS plot of Fig~\ref{Fig: symmetric IPR E DOS}(c). In contrast to this, in the case of asymmetric hoppings, the energy spectrum always remains complex across the DL transition as can be seen from Fig.~\ref{Fig: asymmetric IPR E DOS}(b) and (c).

            \begin{figure}
                \centering
                \includegraphics[scale=0.54]{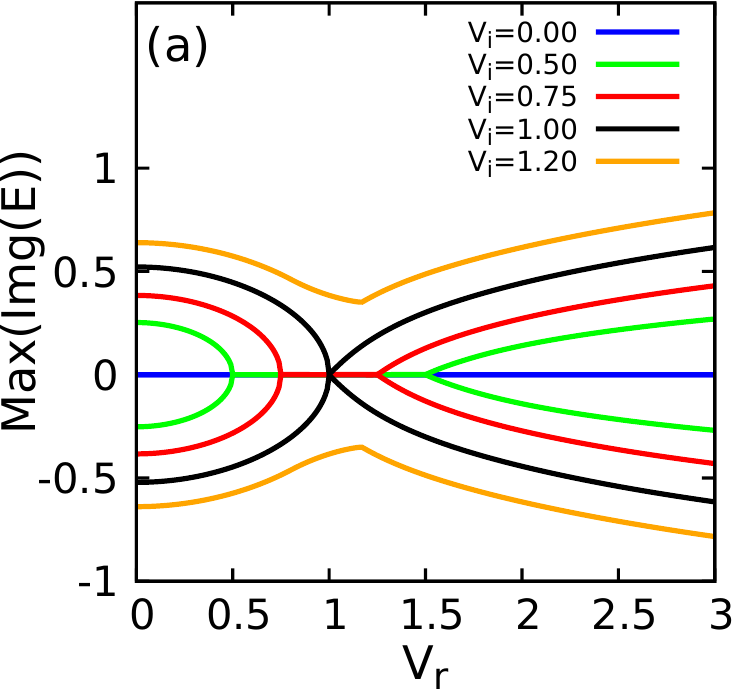}\hspace{2mm}
                \includegraphics[scale=0.54]{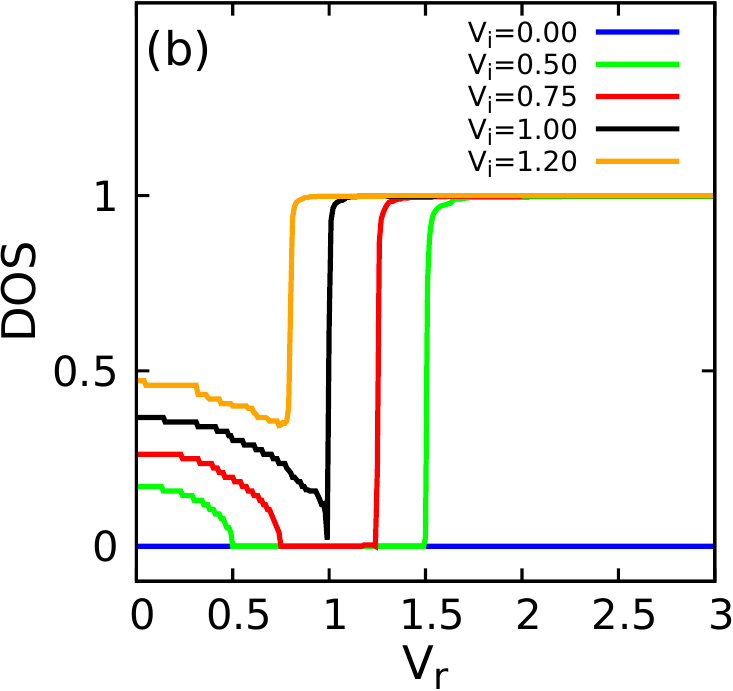}\vspace{2mm}
                \includegraphics[scale=0.54]{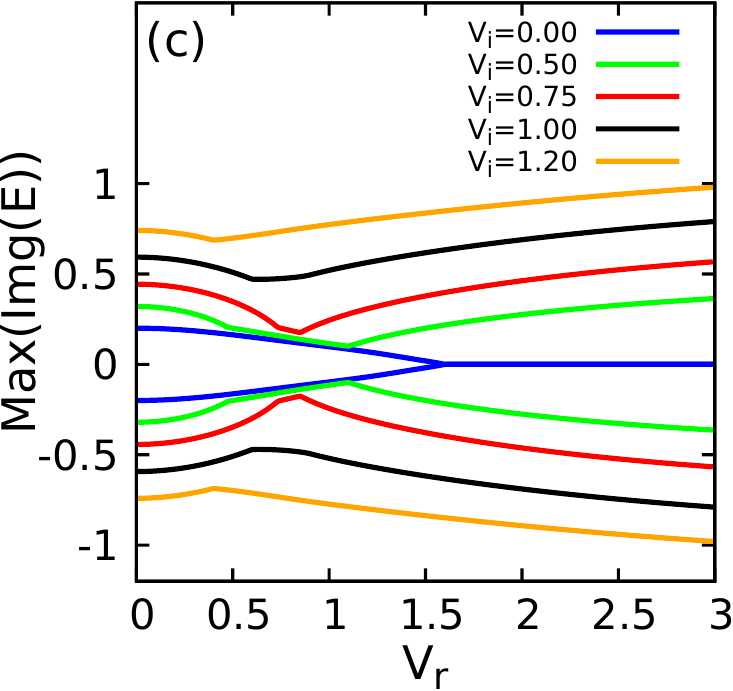}\hspace{2mm}
                \includegraphics[scale=0.54]{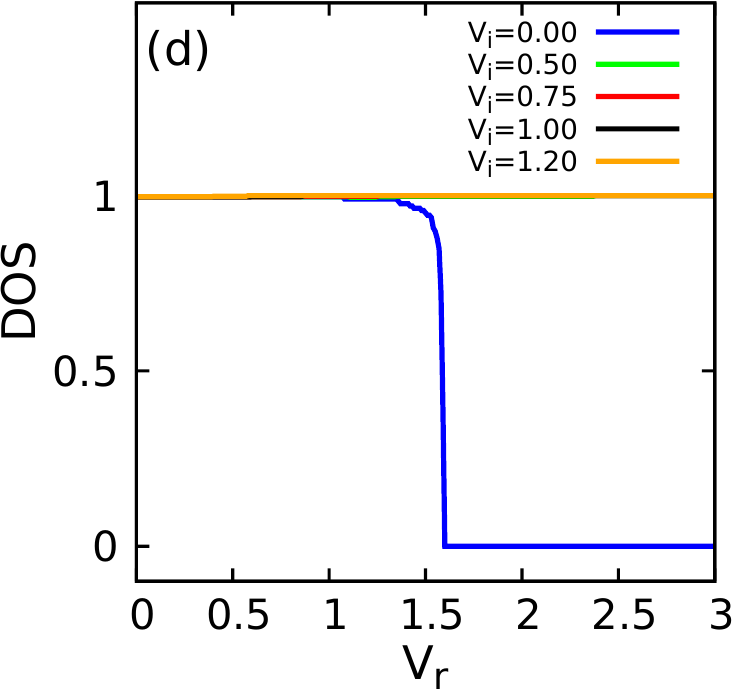}
    
                \caption{The above figures represent a system of size L=610 with an offset phase of $\delta=0.50\pi$. In the upper panel, figures (a) and (b) correspond to systems with symmetric hopping, where $t=t_l=t_r=1.00$. On the other hand, the lower panel, figures (c) and (d), depict systems with asymmetric hopping, specifically $t_l=0.60$ and $t_r=0.80$. The different amplitudes of the imaginary QP potential are shown in the respective figures.}
                \label{Fig: symmetric-asymmetric E DOS}
            \end{figure}
            
            To gain more insight into the changes happening in the eigenspectrum when an asymmetry is introduced in the hoppings, we have plotted the max(Img(E)) and DOS in Fig. \ref{Fig: symmetric-asymmetric E DOS} with different strengths of $V_i$. From Fig. \ref{Fig: symmetric-asymmetric E DOS}(a), we can observe that in the case of symmetric hoppings, as long as $V_i < 1.0$, the eigenspectrum can be purely real. 
            For $V_i \ge 1.0$, the spectrum is no longer purely real for any value of 
            $V_r$. This change in the spectrum  is clearly visible from the DOS plot presented in Fig.~\ref{Fig: symmetric-asymmetric E DOS}(b). It clearly shows that for $V_i < 1.0$, the spectrum is either partly complex or purely real below the critical point (delocalized regime), while it is purely complex above the critical point (localized regime). However, for $V_i \ge 1.0$, there is a transition from partly complex to a fully complex eigenspectrum.
            
            In sharp contrast to the symmetric hopping case, irrespective of the strength of the complex potential $V_i$ (finite value), the spectrum always remains \emph{purely} complex irrespective of delocalization or localization as can be seen in Figs.~\ref{Fig: symmetric-asymmetric E DOS}(c) and (d). In the particular case when the QP potential is completely real, i.e., $V_i=0$, our Hamiltonian becomes the Hatano-Nelson model with QP potential. In this case, the DL transition is accompanied by a complex to the real transition of the eigenspectrum. The appearance of real eigenvalues is typically associated with the restoration of $\mathcal{T}$ symmetry of the eigenfunctions \cite{Zhang2022} in the localized regime. Following this argument, we can see that in the presence of $V_i$, the $\mathcal{T}$ symmetry never gets restored for the eigenfunctions, which leads to a purely complex eigenspectrum across the critical point.

        \subsection{Skin effect, DL transition and eigenspectrum transition in OBC}\label{Sec: OBC}

            \begin{figure*}
                \centering
                \includegraphics[scale=0.7]{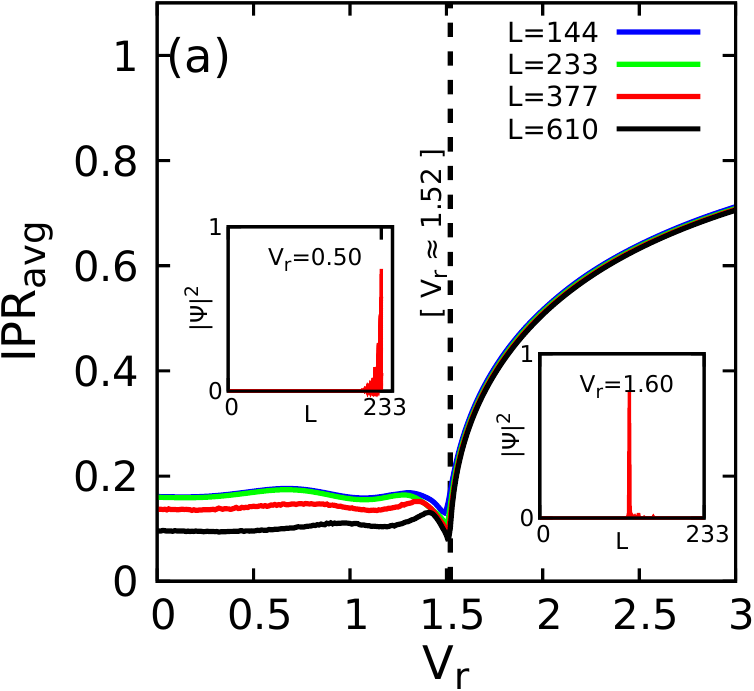}\hspace{8mm}
                \includegraphics[scale=0.7]{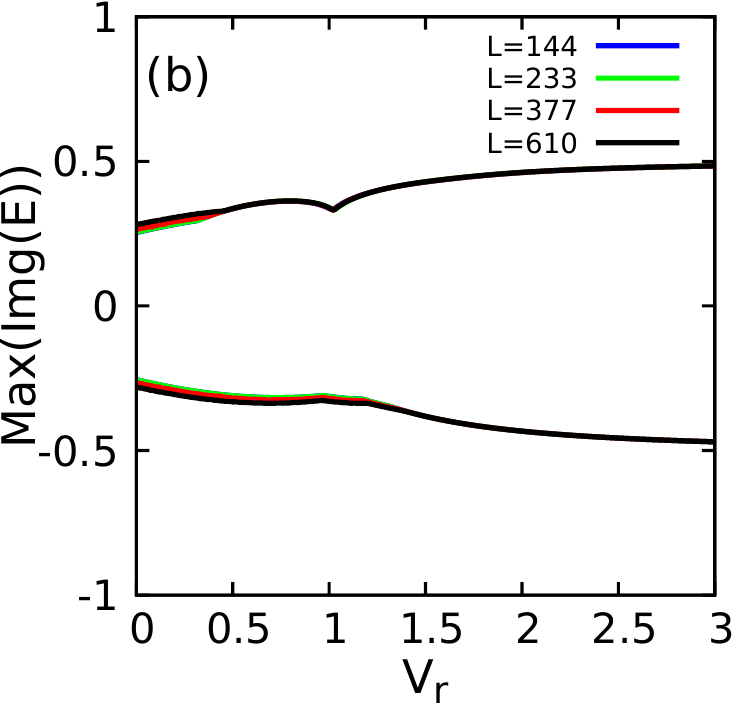}\hspace{8mm}
                \includegraphics[scale=0.7]{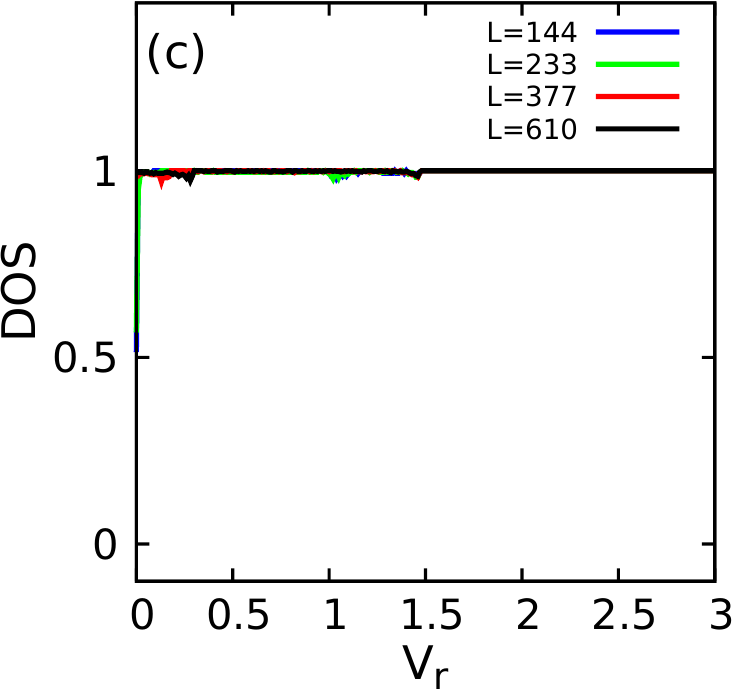}

                \caption{All figures depict the results for a fixed hopping amplitude, with $t_r=0.80$ and $t_l=0.60$, and an offset phase $\delta$ set to zero. The strength of the imaginary QP potential is fixed at $V_i=0.50$. The figures include data for four different system sizes: L=144, 233, 377, and 610, as indicated in the legends. The inset figures within the $IPR_{avg}$ plots illustrate the square of the amplitude of the wave function as a function of system size $L$, demonstrating the transition from the skin states to a localized state.}
                \label{Fig: OBC Asymmetric}
            \end{figure*}
            
            \begin{figure*}
                \centering
                \includegraphics[scale=0.7]{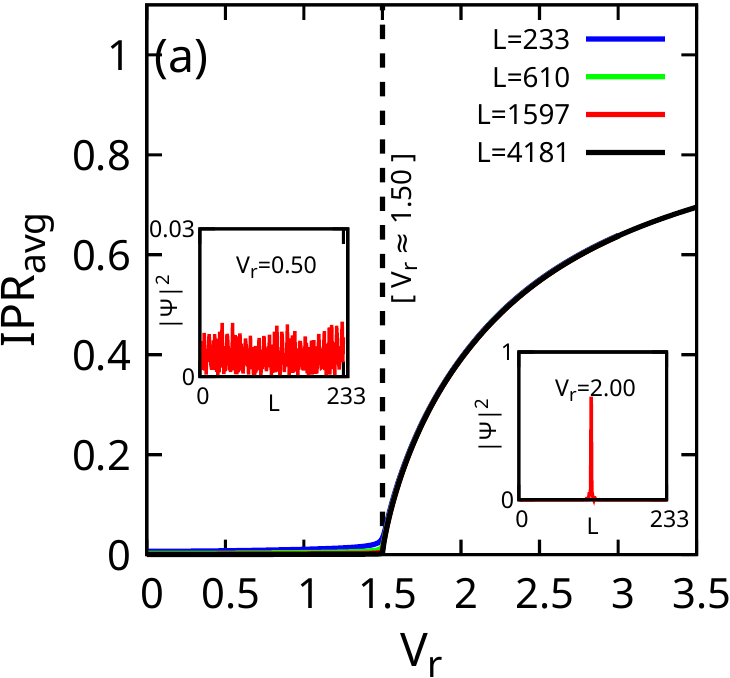}\hspace{8mm}
                \includegraphics[scale=0.7]{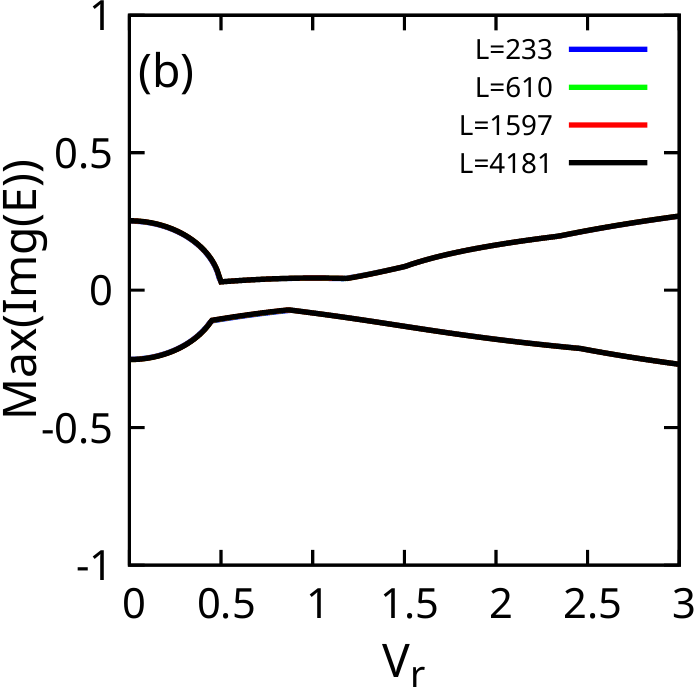}\hspace{8mm}
                \includegraphics[scale=0.7]{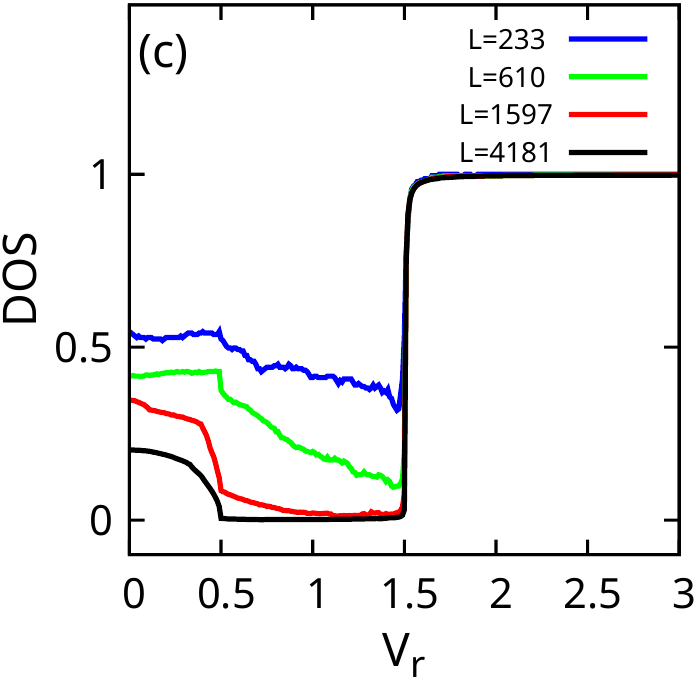}
                
                \caption{The figures presented above correspond to a fixed hopping amplitude, with $t_r=t_l=1.00$, and an offset phase of $\delta=0.50\pi$. The strength of the imaginary QP potential is set to $V_i=0.50$. Four different system sizes, namely L=144, 233, 377, and 610 (as indicated in the figure legends), are plotted. Additionally, the inset figures in the $IPR_{avg}$ plots depict the square of the amplitude of the extended (bulk) wave function as a function of the system size $L$, providing insight into the delocalization-localization transition.}
                \label{Fig: OBC Symmetric}
            \end{figure*}

            It is well known that the nature of the eigenstates crucially depends on the boundary condition in non-Hermitian systems. In particular, under Open Boundary Conditions (OBC) below a certain critical strength of the potential, the asymmetry in the hoppings leads to the accumulation of the eigenstates at one of the boundaries, which is commonly known as the skin effect.  In this section, we have studied the effect of OBC on our model Hamiltonian. As depicted in Fig.~\ref{Fig: OBC Asymmetric}(a) for asymmetric hopping, the critical point remains the same as with PBC, and it agrees with the analytical result in Eq. \ref{Eq: DL critical point}. However, with OBC, below the critical point, we only have skin modes in contrast to the delocalized states in the presence of PBC \cite{Jiang}. Depending on the asymmetry in the hoppings, the skin modes get localized at one of the boundaries, giving rise to a finite $IPR_{avg}$ below the critical point. Since we have $t_r > t_l$, the skin modes appear at the right edge of the lattice, while all the eigenstates are localized in bulk above the critical point. This can be seen in Fig.~\ref{Fig: OBC Asymmetric}(a). The eigenspectrum remains qualitatively similar irrespective of the boundary conditions. From Figs.~\ref{Fig: OBC Asymmetric}(b) and (c), we can observe that, like the PBC case, the spectrum remains purely complex across the critical point in the case of OBC.

            In the case of symmetric hoppings, the nature of the eigenstates below the DL transition strongly depends on system size. The critical point remains unaltered even with OBC, as can be observed from Fig.~\ref{Fig: OBC Symmetric}(a). Below the critical point, the eigenspectrum contains few edge states with complex eigenvalues, which survive even for very large system sizes. On the other hand, with the increase in the system size, the number of extended states with complex eigenvalues decreases, eventually giving rise to an almost identical DOS as in the case of PBC. The eigenspectrum, however, never really becomes purely real owing to the presence of a few edge states. This evolution of the eigenstates with the system size is presented in Fig.~\ref{Fig: OBC Symmetric}(c). Our observation is also corroborated by the data plotted in Fig.~\ref{Fig: OBC Symmetric}(b), which clearly shows that a completely real eigenspectrum never appears with OBC.  

        \subsection{Appearance of mixed state with different frequencies in QP potentials}\label{Sec: Different Freq}

\begin{figure}
                
                \includegraphics[scale=0.5]{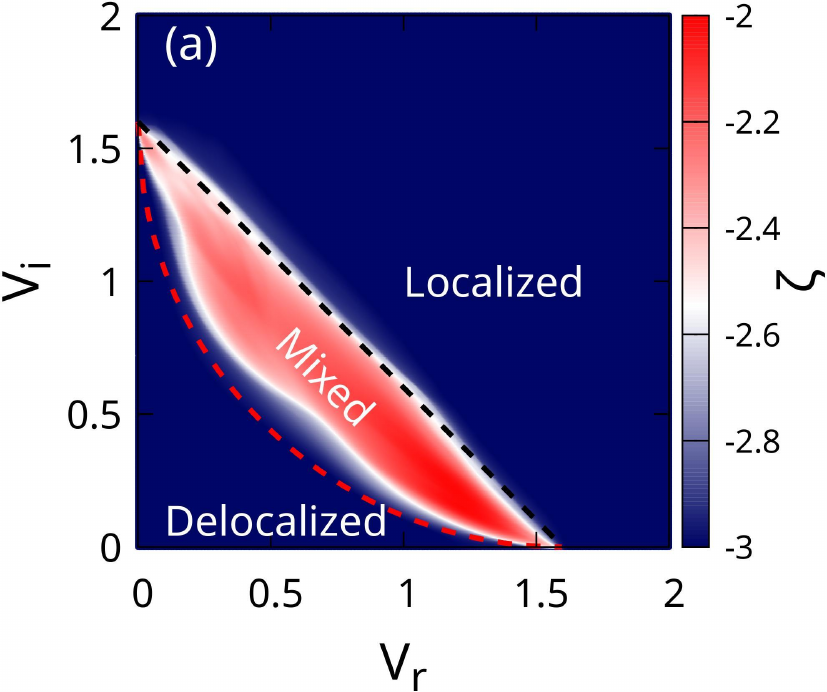}\vspace{3mm}
                \includegraphics[scale=0.5]{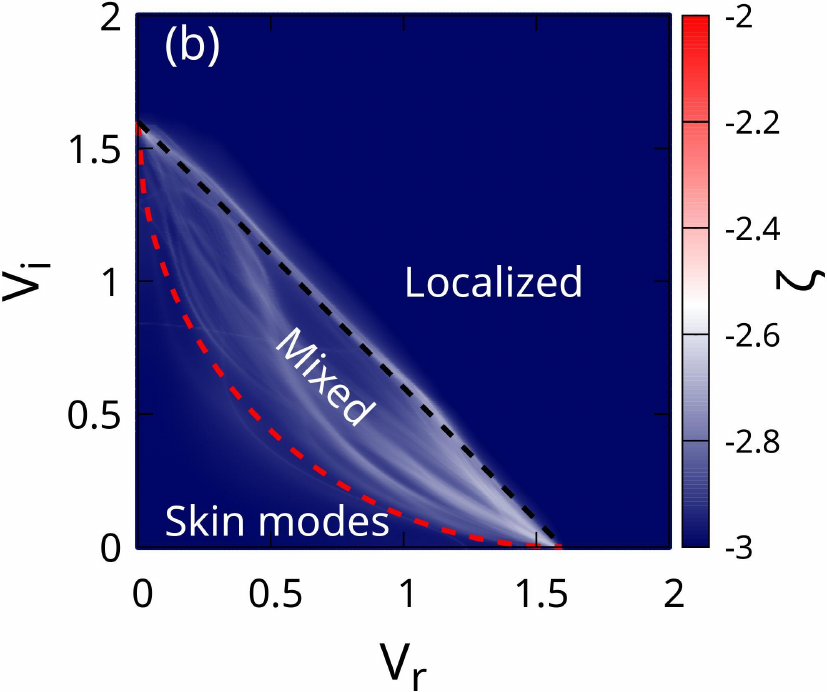}

                \caption{The figures depict phase diagrams for a system with different boundary conditions. Figure (a) shows the phase diagram for PBC, while Figure (b) corresponds to OBC. The system size is $L=1189$, with left and right hopping strengths of $t_l=0.60$ and $t_r=0.80$, respectively, and an offset phase of $\delta=0$. The crucial parameters $\beta_r$ set to 610/989 (golden mean - $\beta_g$ ) and $\beta_i$ set to 360/1189 (bronze mean - $\beta_b$ ). The color scheme represents different regions or phases in the phase diagrams.
                }
                \label{Fig: different freq}
            \end{figure}

            In this section, we further extend our numerical study of our model Hamiltonian (Eq. \ref{Eq: Dual_QP_Hamiltonian}) by considering different incommensurate modulating frequencies ($\beta$) for the real and complex QP potentials. The potential (Eq. \ref{Eq: Quasi Potential}) is now expressed as,
            \begin{eqnarray}
                 V_n = V_r \cos(2\pi{\beta_r}\textit{n} + \phi) + i V_i \cos(2\pi{\beta_i}\textit{n} + \phi +\delta),
                 \label{Eq: Quasi Potential_different_beta}
            \end{eqnarray}
            where $\beta_r$ and $\beta_i$ represent the modulating frequencies corresponding to the real and imaginary QP potentials, respectively. For our purpose, we have chosen $\beta_r = (\sqrt{5}-1)/2$, the golden mean, and $\beta_i= (\sqrt{13}-3)/2$, the bronze mean. 
            From the phase diagram presented earlier, we found that when the real and complex QP potentials have the same frequency ($\beta_r = \beta_i = \beta$), there is a sharp transition between the delocalized and localized phases. However, this behavior is not maintained when the frequencies are non-identical. We have found that irrespective of the boundary conditions, an intermediate phase consisting of localized and delocalized states appears in between the purely delocalized and localized phases, as shown in Figure \ref{Fig: different freq}. To locate the boundaries between different phases, we numerically estimate the parameter $\zeta$, given by
            \begin{equation}
                \zeta=log(IPR_{avg}*NPR_{avg}).
            \end{equation}
            Here, $\rm{IPR_{avg}}$ is given by Eq.~\ref{Eq: IPR_avg}, whereas its complementary function Normalized Participation Ratio ($\rm {NPR_{avg}}$) is defined as $\frac{1}{L^2} \sum_i (IPR^i)^{-1} $. In the purely delocalized region, $\rm{IPR_{avg}}$ tends to zero, while $\rm{NPR_{avg}}$ has a finite value, and vice versa in the purely localized region. For the case of PBC, both the purely delocalized and purely localized regions exhibit a similar numerical value of $\zeta$, represented by the \emph{dark-blue} color in Figure \ref{Fig: different freq}(a). However, in the mixed region where a combination of delocalized and localized states coexists, both $\rm{IPR_{avg}}$ and $\rm{NPR_{avg}}$ have finite values. Therefore, the numerical value $\zeta$ turns out to be less negative in that region, which is indicated by the \emph{red} color in the same phase diagram. While considering OBC, all the purely delocalized states in the case of PBC are replaced by the skin modes as a consequence of the asymmetry in hoppings. This leads to a larger value of $\rm{IPR_{avg}}$ (refer Fig.\ref{Fig: OBC Asymmetric}-a) in contrast to the purely delocalized states for a given system size. However, $\rm{IPR_{avg}}$ for skin modes depends inversely on the system size quite strongly. For purely localized states, $\rm{IPR_{avg}}$ values are identical irrespective of the boundary conditions. This leads to a similar numerical value of $\zeta$ in the case of pure skin modes as well as the localized states, which can be identified by similar dark-blue color in the phase diagram presented in Fig.~\ref{Fig: different freq}(b). Interestingly, from Fig.~\ref{Fig: different freq}(b), we can observe a mixed phase of skin modes and localized states in between the pure skin modes and the localized regime. Furthermore, it is interesting to note that the phase boundary between the mixed phase and the localized regime still follows the equation $V_r+V_i=2t'e^\eta$, which is Eq.~\ref{Eq: Critical} with $\delta=0.$ The other phase boundary that separates the purely delocalized states (or skin modes in the case of OBC) from the mixed state is approximately described by the equation $[V_r-(2t'e^\eta)]^2+[V_i-(2t'e^\eta)]^2=(2t'e^\eta)^2$.
            
    \section{Conclusion}\label{Sec: Concusion}
            In conclusion, we have studied a generalized non-Hermitian 1D Hamiltonian with asymmetric hoppings and dual QP potentials (real and complex). Our investigations reveal that under PBC, the qualitative nature of the phases and the phase boundary remain unaltered compared to the Hamiltonian with symmetric hoppings. In both cases, there exist delocalized and localized phases separated 
            by a sharp phase boundary. Our analytical calculation of the phase boundary, obtained for identical modulations of the dual QP potentials, reveals quantitative differences between the asymmetric and the symmetric cases. Our conclusions are also corroborated numerically. With OBC, however, there is a qualitative difference between the phases of the asymmetric and symmetric cases. For identical modulation of the dual QP potentials, there is a transition from delocalized to localized states in the case of symmetric hoppings. In contrast to this, in the presence of asymmetry in the hoppings, there is a sharp transition from skin modes to localized states at a critical value of the QP potentials. Interestingly, our numerical results indicate that the phase boundary agrees excellently with the analytical expression obtained for PBC. 

            Furthermore, our investigation reveals fascinating distinctions between the cases of equal and unequal incommensurate frequencies in the dual QP potentials. However, depending on the boundary condition, in the case of unequal frequencies, a mixed state occupies the intermediate region between the distinct phases. This mixed state,  characterized by a coexistence of delocalized or skin modes and localized states, represents a novel behavior that is absent when the frequencies in the QP potentials are equal. These findings showcase the intricate nature of non-Hermitian systems and the crucial role of specific choices of QP potentials and their frequencies in the system's behavior.

    \begin{acknowledgments}
        The computation work is supported by SERB (DST), India (Grant No. EMR/2015/001227).
    \end{acknowledgments}

    \appendix
    \section{Calculation of LE for symmetric hopping case (i.e., \texorpdfstring{$t=t_l=t_r$} ))}\label{App: LE_I}
        From Eq. \ref{Eq: Dual_QP_Hamiltonian}, we could write it down in the form of a transfer matrix as,
        \begin{eqnarray}
            \begin{pmatrix}
                  \phi_{n+1} \\
                  \phi_{n}
            \end{pmatrix}
        = T_n
            \begin{pmatrix}
                \phi_{n} \\
                \phi_{n-1}
            \end{pmatrix},
        \end{eqnarray}
        where transfer matrix
        \begin{eqnarray}
        T_n=
            \begin{pmatrix}
                (V_n-E)/t & -1\\
                1 & 0
            \end{pmatrix}.
        \end{eqnarray}
        For the calculation of LE, the transfer matrix of the entire system is required, which is  $T_L=\Pi^L_{n=1}T_n$. Now single particle LE (inverse localization length) can be written as
        \begin{eqnarray}
            \gamma_1(E)=\lim_{x \to \infty} \frac{1}{L} \ln ||T_L||,
        \end{eqnarray}
        here symbol $||~||$ represents the norm of the matrix, which is defined as the maximum absolute value among its eigenvalues.Thus we set the global phase $\theta \to \theta + i\epsilon$ and frequencies of two QP potentials we could apply the Avila's global theory of single-frequency analytical $SL(2,\mathbb{C})$ cocycle \cite{Avila2015, Avila2017}.
        Now we try to get the transfer matrix $T_n$ with two limits of the $\epsilon$, i.e., with limit $\epsilon \to +\infty$ and $\epsilon \to -\infty$, which leads to direct simplified transfer matrix, which reads as
        \begin{eqnarray}
            \lim\limits_{(\epsilon \to +\infty)} T_n &=e^{-i2\pi\beta n -i\theta + \epsilon}
            \begin{pmatrix}
                \frac{V_r+iV_i e^{-i\delta}}{2t} & 0\\
                 0 & 0
            \end{pmatrix}
            + o(1),\\
            \lim\limits_{(\epsilon \to - \infty)}T_n &=e^{i2\pi\beta n +i\theta - \epsilon}
            \begin{pmatrix}
                \frac{V_r+iV_i e^{i\delta}}{2t} & 0\\
                0 & 0
            \end{pmatrix}
            + o(1),~~~
        \end{eqnarray}
        Which leads to LE,
        \begin{eqnarray}
            \gamma_1(E,\epsilon \to +\infty) = +\epsilon + \ln \bigg\vert \frac{V_r+iV_i e^{-i\delta})}{2t} \bigg\vert, \\
            \gamma_1(E,\epsilon \to -\infty) = -\epsilon + \ln \bigg\vert \frac{V_r+iV_i e^{+i\delta})}{2t} \bigg\vert.
        \end{eqnarray}
        According to Avila's global theory, the function $\gamma(E,\epsilon)$ exhibits a convex shape and consists of piecewise linear segments with integer slopes. This states
        \begin{align}
            \gamma_1 &= max \left\{\ln \bigg\vert\frac{V_r + i V_i e^{-i\delta}}{2t}\bigg\vert,\ln \bigg\vert\frac{V_r + i V_i e^{i\delta}}{2t}\bigg\vert,0\right\}\\
            &=max \left\{\frac{1}{2} \ln \frac{V_r^2 + V_i^2 + 2 V_i V_r|\sin \delta|}{(2t)^2},0\right\} \nonumber \\
            &=max \left\{\ln \frac{V_{eff}}{2t},0\right\}
        \end{align}
        For simplify the expression we use $V_{eff}=\sqrt{V_r^2 + V_i^2 + 2 V_i V_r|\sin \delta|}$. It is well known that for $\gamma_1=0$ is the critical point for phase transition and $\gamma_1>0$, all states are fully localized, and for $\gamma_1<0$, all states are delocalized. So DL transition point is determined by the condition
        \begin{eqnarray}
            V_r^2 + V_i^2 + 2 V_i V_r|\sin \delta|=(2t)^2.
        \end{eqnarray}
        LE does not depend on energy, indicating the absence of a mobility edge. All energy states exhibit similar localization properties. For $\delta=\pm\pi/2$ this relation reduces to,
        \begin{eqnarray}
            V_r + V_i = 2t
        \end{eqnarray}
    
    \bibliography{reference}
\end{document}